\documentclass{article}
\usepackage{spconf,amsmath,graphicx,amssymb}
\usepackage{array}

\usepackage{hyperref}
\usepackage{wrapfig}
\usepackage{graphicx}
\usepackage{subfigure}
\usepackage{float}
\usepackage{adjustbox}
\usepackage[justification=centering]{caption}
\usepackage{tikz}
\usepackage{tabularx}
\usepackage{multirow}
\usepackage[belowskip=-15pt,aboveskip=0pt]{caption}

\newcolumntype{P}[1]{>{\centering\arraybackslash}p{#1}}
\newcolumntype{M}[1]{>{\centering\arraybackslash}m{#1}}
\newcommand{\PreserveBackslash}[1]{\let\temp=\\#1\let\\=\temp}
\newcolumntype{R}[1]{>{\PreserveBackslash\raggedleft}p{#1}}
\usepackage{tabulary,booktabs}
\usepackage{enumitem}
\usepackage{lipsum}
\usepackage{blindtext}
\usepackage[linesnumbered, algo2e, ruled,norelsize]{algorithm2e}
\usepackage{xcolor}
\usepackage{cite}

\SetCommentSty{mycommfont}
\title{ End-to-end learned Lossy Dynamic Point Cloud Attribute Compression}
\name{Dat Thanh Nguyen$^\dagger$, Daniel Zieger$^{\ast}$, Marc Stamminger$^{\ast}$, Andr\'e Kaup$^\dagger$ \thanks{This work was funded by the Deutsche Forschungsgemeinschaft (DFG, German Research Foundation) under Grant SFB 1483 – Project-ID 442419336.}}
\address{$^\dagger$Chair of Multimedia Communications and Signal Processing, $^{\ast}$Chair of Visual Computing\\ Friedrich-Alexander-Universität Erlangen-Nürnberg (FAU) \\ Erlangen, Germany}

\begin{document}
%
\maketitle
\begin{abstract}
Recent advancements in point cloud compression have primarily emphasized geometry compression while comparatively fewer efforts have been dedicated to attribute compression. This study introduces an end-to-end learned dynamic lossy attribute coding approach, utilizing an efficient high-dimensional convolution to capture extensive inter-point dependencies. This enables the efficient projection of attribute features into latent variables. Subsequently, we employ a context model that leverage previous latent space in conjunction with an auto-regressive context model for encoding the latent tensor into a bitstream. Evaluation of our method on widely utilized point cloud datasets from the MPEG and Microsoft demonstrates its superior performance compared to the core attribute compression module Region-Adaptive Hierarchical Transform method from MPEG Geometry Point Cloud Compression with 38.1$\%$ Bjontegaard Delta-rate saving in average while ensuring a low-complexity encoding/decoding. 
\end{abstract}
\begin{keywords}
Dynamic Point Cloud, Deep Learning, RAHT, G-PCC.
\end{keywords}
\section{Introduction}
\label{sec:intro}
A point cloud comprises spatially coordinated 3D points with associated attributes, such as color and velocity. Storing millions of 3D points demands substantial storage, especially for dense point clouds where color components require a significantly higher bitrate than geometry components. Consequently, there is a pressing need for effective methods to compress attributes in point clouds. Therefore, our focus in this paper centers on dynamic attribute compression specifically for dense point clouds.

The Moving Picture Expert Group (MPEG) has developed two approaches for Point Cloud Compression (PCC): Geometry-based PCC (G-PCC) and Video-based PCC (V-PCC) \cite{graziosi2020overview}. G-PCC directly compresses 3D point clouds, while V-PCC projects point clouds onto 2D planes, leveraging advancements in image and video compression. Our study assumes lossless geometry encoding, concentrating solely on point cloud attribute compression.
\begin{figure}
\centering
\captionsetup{justification=justified}
\includegraphics[width=.99\linewidth]{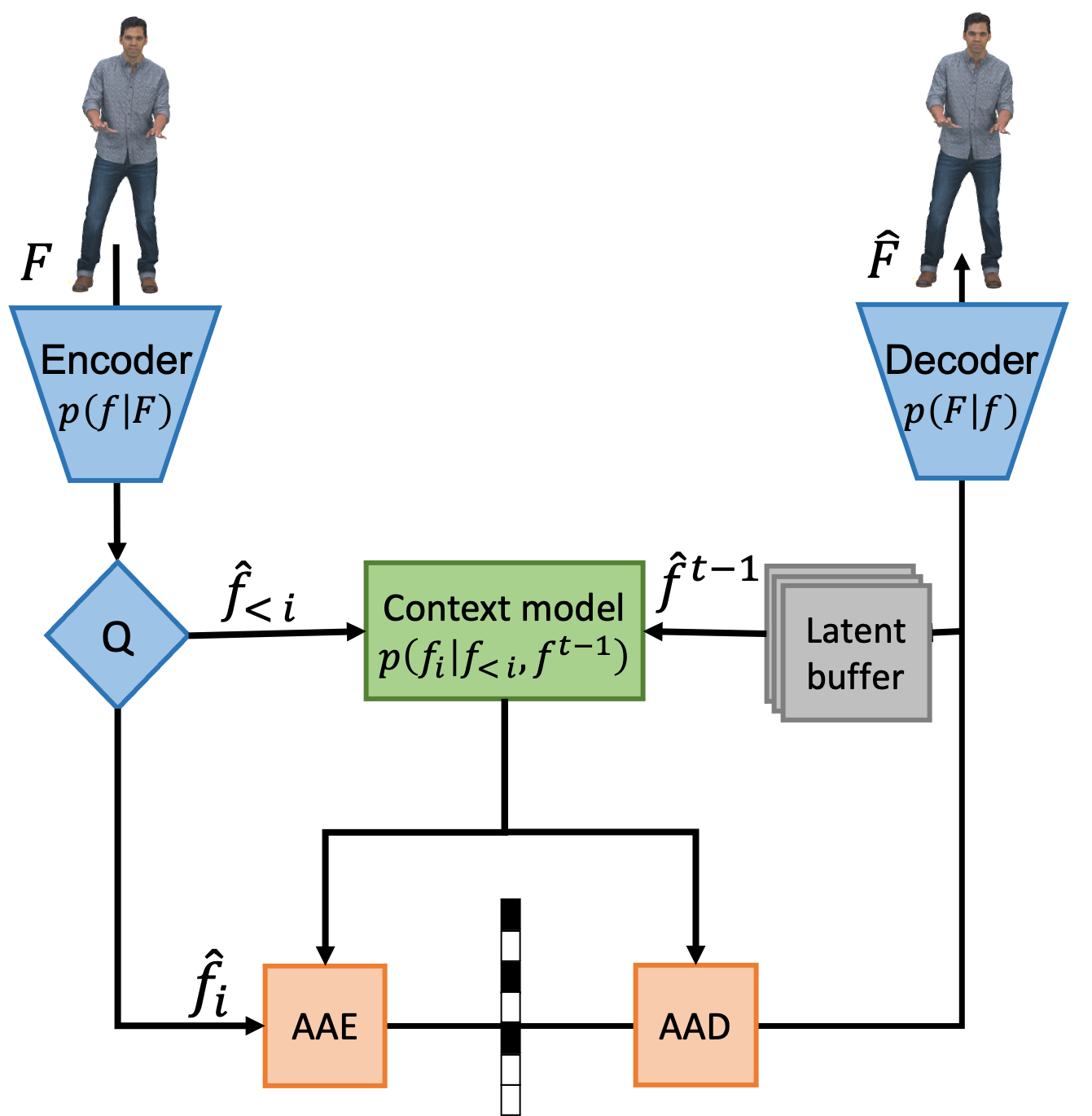}
\vspace{0.1cm}
\caption{System overview of the proposed method. The Encoder encodes each frame into a latent variable $f$ before a quantization step. The quantized latent variables are encoded by an Adaptive Arithmetic Encoder (AAE) using the probability distribution model from the spatiotemporal Context Model. At the decoder side, the quantized latent variables are decoded from the bitstream using the same context model and then fed into the Decoder to reconstruct the lossy feature $\title{F}$.}
\label{fig:3Dcontext}
\end{figure}
Similar to 2D image and video content, exploiting spatial and temporal correlations within point cloud attributes can enhance compression tasks. However, transposing efficient signal processing tools from 2D spaces to sparse and irregularly sampled 3D point data presents challenges. Existing attribute compression approaches mostly rely on geometry-dependent transformations \cite{7025414,7482691,gu20193d}, where the geometry representation significantly influences attribute compression. Point cloud geometry can adopt point, octree, or voxel representations. Voxel representation allows conventional operations like 3DCNN and DCT, albeit at the cost of considerable computational power. Octree representation offers a lower uncompressed bitrate than voxels, but compatibility with conventional operations is limited. In our study, we opt for point-based representation incorporated with an autoencoder-based network (AE) \cite{kingma2013auto} to efficiently compress point clouds.


In contrast to conventional 2D video compression techniques, our approach involves the substitution of explicit motion prediction and residual compensation with an autoencoder-based temporal context model. Theoretically, autoencoders provide an excellent framework for the lossy image/video compression problem \cite{habibian2019video, yang2020learning}, as they offer a mechanism for minimizing reconstruction error and bitrate, measurable through two loss terms. In our investigation, the encoder learns to capture and store important features to be transmitted to the latent variables. We then build a spatiotemporal probability distribution model to model the distributions of the latent variables for encoding. Experimental results demonstrate that our approach outperforms recently proposed methods in attribute compression and rivals the latest G-PCC version while exhibiting superior computational efficiency. 
\par Our two main contributions of this work are listed as follows: $i$) We propose one of the first VAE-based end-to-end learned dynamic point cloud attribute compression frameworks that outperforms the core module of the state-of-the-art PCC algorithm. $ii$) Development of a low-complexity and efficient codec that can generalize for multimodal compression. The remainder of the paper is structured as follows: Section II reviews related work; the proposed method is described in Section III; Section IV presents the experimental results, and conclusions are drawn in Section V.
\section{Related work}
\label{sec:stateoftheart}

Many methods for compressing point cloud attributes rely on transforms that depend on geometry \cite{7025414,7482691,gu20193d}. For example, the approach proposed in \cite{7025414} constructs a graph from geometry, treating attributes as signals on this graph. The attributes are then transformed using a Graph Fourier Transform (GFT) before undergoing quantization and entropy coding. The Region-Adaptive Hierarchical Transform (RAHT) \cite{de2016compression} transforms the attributes of occupied octree nodes, resembling an adaptive variation of a Haar wavelet. The quantized transform coefficients are subsequently encoded using Run-Length Golomb-Rice encoding (RLGR) with a Laplace distribution assumption. RAHT serves as the core attribute compression module in the MPEG G-PCC TMC13 reference software \cite{TMC13}, alongside the lifting transform \cite{lifting} and the predicting transform \cite{predicting}. However, this method is specifically designed for color attribute compression and when other attributes are includes, RAHT might fail to exploit dependencies between various modalities. In this work, we also focus on color attribute compression, but our method can be easily adapted to other point cloud attributes such as velocities, normals, etc.

\par One direction for point cloud attribute coding involves projecting 3D space onto 2D planes and utilizing existing image and video compression standards (e.g., JPEG and H.265). For instance, the method presented in \cite{7434610} partitions the point cloud into 8x8 blocks, utilizes snake scanning, converts the data to 2D images, and encodes it using the JPEG standard. MPEG V-PCC \cite{graziosi2020overview} projects point cloud data into three associated videos: occupancy map, geometry video, and attribute video. All three generated videos are then encoded using 2D video coding standards.

\par Recently, several studies \cite{quach2020folding,Sheng2021DeepPCACAE,wang2022sparse,10095385} have proposed applying advanced deep learning techniques to lossy point cloud attribute compression. In \cite{quach2020folding}, a folding network is introduced to project 3D attribute signals to 2D planes. The authors in \cite{wang2022sparse} propose an autoencoder neural network to map attributes to a latent space, followed by encoding the latent vector using a hyperprior/auto-regressive context model. In this work, point clouds are represented using sparse tensors, enabling efficient processing through sparse convolution. The autoencoder architecture is also employed in the static lossy attribute coding proposed in \cite{9447226}. Drawing inspiration from recent works in 2D space \cite{habibian2019video,yang2020learning} utilizing autoencoders and spatiotemporal context models in video compression, we establish a variational autoencoder-based architecture for lossy dynamic 3D point cloud attribute coding.

\par There has been more emphasis on lossy intra-frame coding than inter-frame coding, resulting in a paucity of works on dynamic point cloud attribute coding. In MPEG V-PCC \cite{graziosi2020overview}, point clouds are projected to 2D space, and spatial and temporal correlations between frames are exploited using conventional 2D transformations or motion prediction. In \cite{fang20224dac}, authors identify the nearest neighbor in the reference frame and use its attributes for prediction. They then construct a 3D attribute embedding network to estimate the target attributes and obtain attribute residuals. The temporal attribute residuals are subsequently encoded using the RAHT codec \cite{graziosi2020overview}. 
\par In this paper, we establish a framework for dynamic lossy attribute coding with an important assumption that all point cloud attribute data is generated as a continuous function of some latent variables.  Then our goal is to learn the latent variable model (encoder) so we can reconstruct/generate new data (decoder) from a latent variable. To the best of our knowledge, this represents one of the pioneering end-to-end learning-based dynamic point cloud attribute compression approaches.

\begin{figure*}
\centering
\captionsetup{justification=justified}
\includegraphics[width=.96\linewidth]{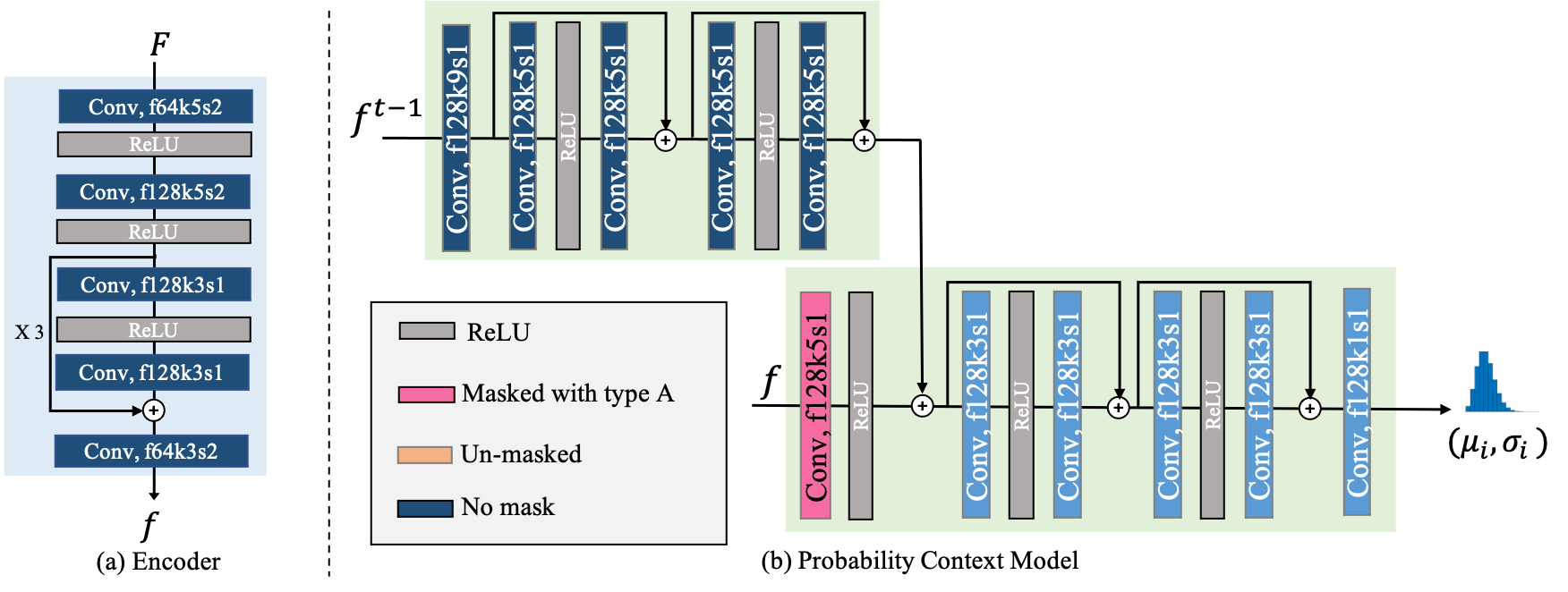}
\vspace{0.1cm}
\caption{Detail the network architecture of the proposed method. $Conv$ denotes sparse convolution \cite{choy20194d}, with the number of filters, convolutional kernel, and stride denoted using $fLkKsS$ (where $L$ represents filters, $K$ represents kernel, and $S$ represents stride). The Probability Context Model outputs a pair of mean $\mu$ and standard deviation $\sigma$ of the Gaussian distribution. The architecture of the Decoder mirrors the Encoder with sparse transpose convolution. }
\label{fig:network}
\end{figure*}

\section{Proposed method}
\label{proposedmethod}
\subsection{Auto-Encoder and Rate-Distortion Optimization}

In this section, we provide a brief summary of the preliminaries and introduce the details of our method. We represent the attributes of each point cloud frame using sparse tensors \{C, F\}, given that the coordinates of occupied voxels $C=\{x_i,y_i,z_i\}_i$ are already known. Our goal is to encode each frame's feature set $F$, composed of red, green, and blue (RGB) colors, or other attributes such as YUV colors, normals, velocities, etc. In our strategy for lossy encoding, we aim to acquire a latent variable model where the latent variable $f$ encapsulates crucial features to be transmitted and subsequently reconstructed by the decoder. In other words, we indirectly model the data distribution by a generative model and an unobserved latent variable. The joint model of the feature $F$ and discrete latent variable $f$ can be written as:

\begin{equation} \label{eq:factorization}
\begin{split}
   p_{\theta}(F, f) = p_{\theta}(f) p_{\theta}(F|f) 
\end{split}
\end{equation}

Both terms on the right-hand side are parametrized distributions, where $p_{\theta}(f)$ is the bottleneck context model, and $p_{\theta}(F|f)$ is the decoder. The encoder $q_{\phi}(f|F)$ can be viewed as providing an accurate posterior estimate of the latent variables $f$ from an input $F$. Since the likelihood $\log p_{\theta}(F)$ is intractable, the variational upper bound \cite{kingma2013auto} on the log-likelihood feature $F$ is optimized instead: 

\begin{equation} \label{eq:upperbound}
\begin{split}
   -\log p(F) \leq \textrm{KL}\left[   q(f|F)|p(f) \right] - \textrm{E}_{q}\left[\log p(F|f) \right]
\end{split}
\end{equation}
\begin{table*}[t]
\caption{The BD rate saving and running time of the proposed method compared to RAHT}
\label{tab:compare}
\begin{center}
\begin{tabular}{@{}|l|l|c|c|c|c|c|c|c|@{}}
\hline
\multirow{3}{*}{\textbf{Dataset}}&\multirow{3}{*}{\textbf{Point Cloud}}& \multirow{3}{*}{\textbf{No. frames} }&\multicolumn{1}{c|}{\textbf{BD-Rate}}
&\multicolumn{1}{c|}{\textbf{BD-Quality}}
&\multicolumn{2}{c|}{\textbf{Coding Time}}\\
 \cline{4-7}
& &      
	&\textbf{PSNR-Y}  
	&\textbf{PSNR-Y}
 &\textbf{RAHT} &\textbf{Proposed} \\

\cline{4-7}
 & 
&&(\%) 
&(dB) 
 &(s) &(s) \\
\hline
\multirow{2}{*}{MVUB}&Phil&244&-36.11&1.24&1.4&2.5\\
\cline{2-7}
&Ricardo&299&-29.90&1.44&0.9&1.6\\
\hline
\multirow{2}{*}{MPEG 8i}&Redandblack&299&-9.27&0.28&0.7&1.5\\
\cline{2-7}
&Loot &299&-56.21&2.46&0.7&1.7\\
\hline
\multirow{2}{*}{Owlii}&Basketball&600&-51.42&1.82&0.7&1.2\\
\cline{2-7}
&Dancer&600&-48.40&1.42&0.6&1.1\\
\hline

\multicolumn{3}{|c|}{\textbf{Average}}& \textbf{-38.10}&\textbf{1.44}&\textbf{0.9}&\textbf{1.8}\\
\hline

\end{tabular}
\end{center}
\end{table*}

We use neural networks to parameterize both $q(f|F)$ and $p(F|f)$ (the encoder and decoder). The second term on the right-hand side in Eq.\ref{eq:upperbound} measures the reconstruction error of $F$ given a sample $f \sim q(f|F)$. Please note that unlike our previous work \cite{10024999, 10095385} where we losslessly encode $F$ by feeding $p(F|f)$ to an arithmetic coder, in this work, we only send $f$ to the decoder to reconstruct $F$.  

The first regularization term in Eq. \ref{eq:upperbound} is related to the cost of coding the latent $f$ produced by the encoder $q(f|F)$ using an optimal code derived from the prior $p(f)$. Note that Kullback–Leibler divergence between two distributions $KL[q|p] = CE[q|p] - H[q]$ and $H[q] =0$ for a deterministic encoder $f= \phi(F)$; therefore, the regularization term can be replaced by the cross-entropy between $p$ and $q$: 

\begin{equation} \label{eq:loss}
\begin{split}
   \textrm{KL}[q(f|F)|p(f)] = \textrm{CE}[q(f|F)|p(f)] = -\log p(f)
\end{split}
\end{equation}

Thus, by adding a tradeoff parameter $\lambda$, we can reformulate the rate-distortion loss function in Eq \ref{eq:upperbound} as:

\begin{equation} \label{eq:loss}
\begin{split}
   L = \textrm{E}_{q} \left[ -  \log p(f)     -  \lambda \log p(F|f) \right] 
\end{split}
\end{equation}

We model $p(f)$ by incorporating auto-regressive context and temporal context as: 

\begin{equation} \label{eq:loss}
\begin{split}
   p(f, f^{t-1}) = \prod_{i=1}^{N} p(f_i|f_{<i},f^{t-1})
\end{split}
\end{equation}

where $f_{<i}$ denotes the spatial elements that come before $i$ in the auto-regressive ordering, and $f^{t-1}$ denotes the previous temporal latent variable. Putting them altogether, we have the final loss: 

\begin{equation} \label{eq:loss}
\begin{split}
   L = \textrm{E}_{q}\left[  - \sum_{i=1}^{N} \log p(f_i|f_{<i},f^{t-1}) - \lambda  \log p(F|f) \right]
\end{split}
\end{equation}

\subsection{Network Architecture}
We design a lightweight yet effective auto-encoder architecture for fast encoding/decoding time. At the encoder side, we start with two convolution layers with stride=2 to downscale the input point clouds, each followed by a $ReLU$ activation layer. Next, we deploy three residual blocks before down-scaling again with a convolutional layer with 64 filters. The encoder outputs continuous latent variables $f$, which are then quantized. We quantize the latent variable by assigning each element $z_i$ to the closest integer values; however, this rounding is only performed in the inference. To allow gradient backpropagation during the training, we replace the rounding function by an additive uniform noise: 

\begin{equation} \label{eq:loss}
\begin{split}
    \Tilde{f}=f+u, u\sim U(\frac{1}{2},\frac{1}{2})
\end{split}
\end{equation}

To encode $\Tilde{f}$ into a bitstream, we use a context-adaptive arithmetic encoder (AE) assisted by a probability context model $p(\Tilde{f})$. The probability distribution of each latent variable element $p(\Tilde{f_i})$ is parameterized as a Gaussian distribution with mean $\mu_i$ and standard deviation $\sigma_i$. The probability context model network takes two inputs: the current latent variable $\Tilde{f}$ providing auto-regressive context and the $\Tilde{f}^{t-1}$ for the temporal context. We apply the masked convolution on the auto-regressive path to preserve the causality constraint. The decoder symmetrizes the architecture of the encoder with the final convolution layer with filter size 3 corresponding to the 3 color attributes. To measure the distortion, we simply use the Mean-Square-Error (MSE) loss between the reconstructed point cloud and the input point cloud attribute. 

\subsection{Encoding and Decoding}
To encode a chunk of point cloud frames $F$, we map it through the encoder to obtain the latents $f$. Then we encode the latent variables into a bitstream element by element sequentially. Specifically, every time an element is encoded, it is fed back to the probability context model to predict the probability distribution of the next element $p(f_i|f_{<i}, f^{t-1})$. We then pass the probability distribution to an arithmetic coder to encode the next symbol. 
For decoding, we process the bitstream $b_i$ and combine it with the prediction $p(f_i|f_{<i}, f^{t-1})$ to obtain $f_i$. Finally, we pass all decoded latent variables to the decoder to obtain the reconstructed attributes.

\section{Experimental Results}
\label{performanceeval}
\subsection{Experimental Setup}
\textbf{Training Dataset:} There is only a limited training set for dynamic point cloud attribute coding. We consider widely used point cloud datasets with color attributes, namely MPEG 8i \cite{8i} - dynamic dense voxelized full-body point clouds, and Microsoft Voxelized Upper Bodies (MVUB) \cite{loop2016microsoft} - dynamic dense voxelized upper-body point clouds. We select $longdress$ and $soldier$ from the MPEG 8i dataset and $andrew, david, sarah$ from the MVUB dataset for training.

\textbf{Training Procedure:} The training and testing processes were performed on a GeForce RTX 3090 GPU with an Adam optimizer. Applying an initial learning rate of $10^{-3}$ with a step decay by a factor of 0.95 every 3 epochs yielded optimal results. We used a batch size of 4 and early stopping with a patience of 20 epochs. To obtain rate-distortion curves, we trained separate models for lambda values $\lambda \in$ $\left[0.003, 0.01, 0.1, 0.5, 1.0, 2.5\right] $

\textbf{Test Point Clouds:} In this work, we aim at dynamic dense point cloud coding. We evaluate our methods on a set of dynamic dense point clouds from the MVUB and MPEG datasets. A We test our method on six sequences: $redandblack, loot$ (MPEG 8i); $phil, ricardo$ (MVUB); $basketball, dancer$ (Owlii \cite{owlii}). In total, we have 2257 frames for testing.

\subsection{Experimental Results}
\begin{figure}
\captionsetup{singlelinecheck = false, justification=justified, font=small, labelsep=space}
\begin{minipage}[b]{.99\linewidth}
  \centering
  \centerline{\includegraphics[width=0.8\linewidth]{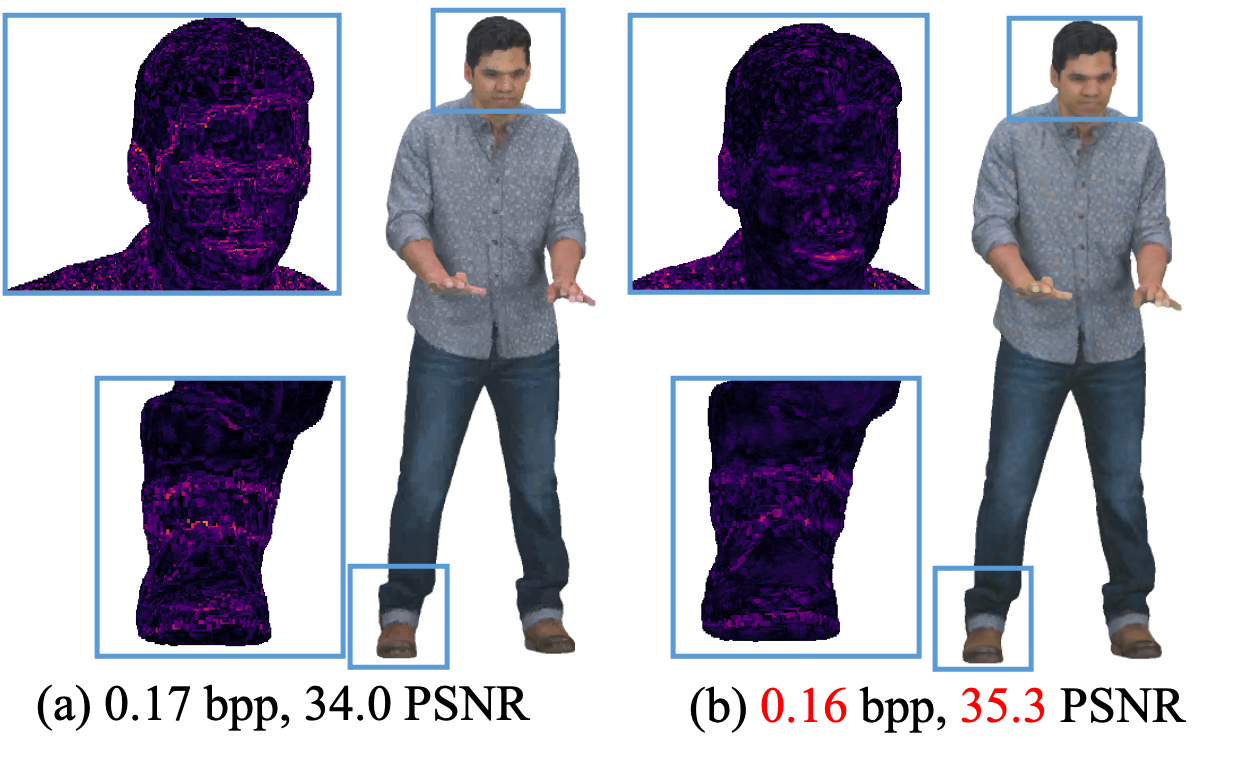}}
\end{minipage}
\hfill
\begin{minipage}[b]{0.99\linewidth}
\label{sfig:typeA}
  \centering
  \centerline{\includegraphics[width=0.9\linewidth]{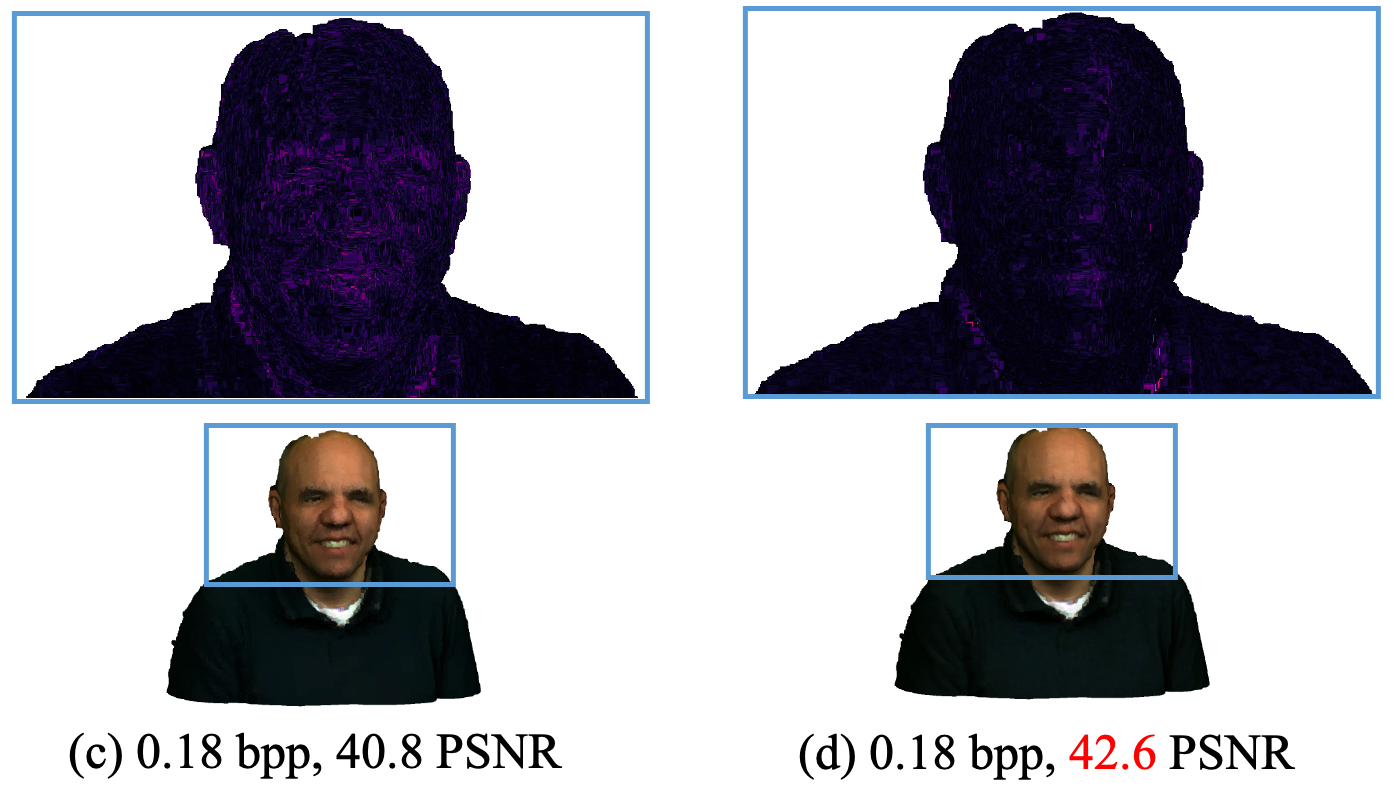} }
\end{minipage}
\hfill

\caption{Visual comparison between RAHT ((a) and (c)) and the proposed method ((b) and (d)) at a similar bitrate of point cloud $Loot$ and $Ricardo$. The darker distortion map is, the better quality. Please note that for better visualization, we lower the max value of the color map to emphasize the distortions.}
\label{fig:distortionvisualization}
\end{figure}

Due to the limited works on point cloud attribute compression, especially source code unavailability of learning-based methods, we benchmark our method against the core attribute compression model from G-PCC \cite{graziosi2020overview} as a widely accepted anchor. For RAHT, we set 8 different $QP$ values $QP \in$ [51, 46, 40, 34, 28, 22, 18, 14]. Table \ref{tab:compare} reports the Bjontegaard-Delta rate using both PSNR-Y and PSNR-YUV results and the encoding time of the proposed method compared with RAHT. We observe that the proposed method consistently outperforms RAHT in both BD-rate and BD-quality metrics. The proposed method provides a significant bitrate reduction of 38.1$\%$ in average at the same quality level. We provide roughly a 1.44dB gain in BD-quality at the same bitrate compared with RAHT. Besides, as we aim to build a low-complexity encoder, the run time is comparable to RAHT. Please note that RAHT is implemented in C++, while the bottleneck of our run time is coming from writing the latent space to bitstream, which is implemented in Python. In Figure \ref{fig:distortionvisualization}, we show the reconstructed attributes from RAHT and the proposed method with a corresponding distortion map. We choose L2Norm between the reconstructed attribute and the input attribute to measure the distortion, the darker color in the distortion map indicates better quality. We observe that the proposed method generally obtains lower distortion across the whole point cloud, especially in regions with geometric details (in the zooming-in regions), while RAHT is unable to predict the color attribute correctly.
%

\section{Conclusions}
\label{conclusion}
We have presented an end-to-end learning framework for dynamic point cloud attribute compression based on a variational autoencoder. The encoder captures important features and maps them to the latent variables before transmitting them to the decoder. We efficiently encode the latent variables by leveraging temporal dependencies among sequential latent elements, incorporating spatial auto-regressive context. We tested the method on a widely used set of datasets and have shown that our method outperforms the core attribute compression module from MPEG G-PCC with low computational complexity. We are now working on improving the coding performance and extending our work to encode the other multimodal attibute data of point clouds.



\bibliographystyle{IEEEbib}
\bibliography{references/IEEEabrv.bib, references/refs.bib}

\begin{thebibliography}{10}

\bibitem{graziosi2020overview}
D.~Graziosi, O.~Nakagami, S.~Kuma, A.~Zaghetto, T.~Suzuki, and A.~Tabatabai,
\newblock ``An overview of ongoing point cloud compression standardization activities: video-based ({V}-{PCC}) and geometry-based ({G}-{PCC}),''
\newblock {\em APSIPA Transactions on Signal and Information Processing}, vol. 9, 2020.

\bibitem{7025414}
C.~Zhang, D.~Florêncio, and C.~Loop,
\newblock ``Point cloud attribute compression with graph transform,''
\newblock in {\em Proc. IEEE International Conference on Image Processing (ICIP)}, 2014, pp. 2066--2070.

\bibitem{7482691}
R.~L. de~Queiroz and P.~A. Chou,
\newblock ``Compression of 3d point clouds using a region-adaptive hierarchical transform,''
\newblock {\em IEEE Transactions on Image Processing}, vol. 25, no. 8, pp. 3947--3956, 2016.

\bibitem{gu20193d}
S.~Gu, J.~Hou, H.~Zeng, H.~Yuan, and K.-K. Ma,
\newblock ``3d point cloud attribute compression using geometry-guided sparse representation,''
\newblock {\em IEEE Transactions on Image Processing}, vol. 29, pp. 796--808, 2019.

\bibitem{kingma2013auto}
D.~P. Kingma and M.~Welling,
\newblock ``Auto-encoding variational bayes,''
\newblock {\em arXiv preprint arXiv:1312.6114}, 2013.

\bibitem{habibian2019video}
A.~Habibian, T.~v. Rozendaal, J.~M. Tomczak, and T.~S. Cohen,
\newblock ``Video compression with rate-distortion autoencoders,''
\newblock in {\em Proceedings of the IEEE/CVF International Conference on Computer Vision}, 2019, pp. 7033--7042.

\bibitem{yang2020learning}
R.~Yang, F.~Mentzer, L.~Van~Gool, and R.~Timofte,
\newblock ``Learning for video compression with recurrent auto-encoder and recurrent probability model,''
\newblock {\em IEEE Journal of Selected Topics in Signal Processing}, vol. 15, no. 2, pp. 388--401, 2020.

\bibitem{de2016compression}
R.~L. De~Queiroz and P.~A. Chou,
\newblock ``Compression of 3d point clouds using a region-adaptive hierarchical transform,''
\newblock {\em IEEE Transactions on Image Processing}, vol. 25, no. 8, pp. 3947--3956, 2016.

\bibitem{TMC13}
M.~Group,
\newblock ``{MPEG} tmc13 reference software,'' (accessed October 25, 2022).

\bibitem{lifting}
K.~Mammou, A.~Tourapis, J.~Kim, F.~Robinet, V.~Valentin, and Y.~Su,
\newblock ``Proposal for improved lossy compression in tmc1,''
\newblock in {\em ISO/IEC JTC1/SC29/WG11 M42640, 2018}.

\bibitem{predicting}
M.~3DG,
\newblock ``{PCC} test model category 3 v0.,''
\newblock in {\em ISO/IEC JTC1/SC29/ WG11 N17249, 2017.}

\bibitem{7434610}
R.~{Mekuria}, K.~{Blom}, and P.~{Cesar},
\newblock ``Design, implementation, and evaluation of a point cloud codec for tele-immersive video,''
\newblock {\em IEEE Transactions on Circuits and Systems for Video Technology}, vol. 27, no. 4, pp. 828--842, 2017.

\bibitem{quach2020folding}
M.~Quach, G.~Valenzise, and F.~Dufaux,
\newblock ``Folding-based compression of point cloud attributes,''
\newblock in {\em Proc. IEEE International Conference on Image Processing (ICIP)}. IEEE, 2020, pp. 3309--3313.

\bibitem{Sheng2021DeepPCACAE}
X.~Sheng, L.~Li, D.~Liu, Z.~Xiong, Z.~Li, and F.~Wu,
\newblock ``Deep-pcac: An end-to-end deep lossy compression framework for point cloud attributes,''
\newblock {\em IEEE Transactions on Multimedia}, vol. 24, pp. 2617--2632, 2021.

\bibitem{wang2022sparse}
J.~Wang and Z.~Ma,
\newblock ``Sparse tensor-based point cloud attribute compression,''
\newblock {\em arXiv preprint arXiv:2204.01023}, 2022.

\bibitem{10095385}
D.~T. Nguyen, K.~G. Nambiar, and A.~Kaup,
\newblock ``Deep probabilistic model for lossless scalable point cloud attribute compression,''
\newblock in {\em Proc. IEEE International Conference on Acoustics, Speech and Signal Processing (ICASSP)}, 2023, pp. 1--5.

\bibitem{9447226}
X.~Sheng, L.~Li, D.~Liu, Z.~Xiong, Z.~Li, and F.~Wu,
\newblock ``Deep-pcac: An end-to-end deep lossy compression framework for point cloud attributes,''
\newblock {\em IEEE Transactions on Multimedia}, pp. 1--1, 2021.

\bibitem{fang20224dac}
G.~Fang, Q.~Hu, Y.~Xu, and Y.~Guo,
\newblock ``4dac: Learning attribute compression for dynamic point clouds,''
\newblock {\em arXiv preprint arXiv:2204.11723}, 2022.

\bibitem{choy20194d}
C.~Choy, J.~Gwak, and S.~Savarese,
\newblock ``4d spatio-temporal convnets: Minkowski convolutional neural networks,''
\newblock in {\em Proceedings of the IEEE Conference on Computer Vision and Pattern Recognition}, 2019, pp. 3075--3084.

\bibitem{10024999}
D.~T. Nguyen and A.~Kaup,
\newblock ``Lossless point cloud geometry and attribute compression using a learned conditional probability model,''
\newblock {\em IEEE Transactions on Circuits and Systems for Video Technology}, vol. 33, no. 8, pp. 4337--4348, 2023.

\bibitem{8i}
``Common test conditions for {PCC},''
\newblock in {\em {ISO}/{IEC} {JTC1}/{SC29}/{WG11} {MPEG} output document {N19324}}.

\bibitem{loop2016microsoft}
C.~Loop, Q.~Cai, S.~O. Escolano, and P.~A. Chou,
\newblock ``Microsoft voxelized upper bodies - a voxelized point cloud dataset,''
\newblock in {\em {ISO}/{IEC} {JTC1}/{SC29} {Joint} {WG11}/{WG1} ({MPEG}/{JPEG}) input document m38673/{M72012}}. May 2016.

\bibitem{owlii}
Y.~Xu, Y.~Lu, and Z.~Wen,
\newblock ``Owlii dynamic human mesh sequence dataset,''
\newblock {\em ISO/IEC JTC1/SC29/WG11 m41658}, 2017.

\end{thebibliography}

\end{document}